\documentclass[twocolumn,iop]{emulateapj}

\usepackage{natbib}

\usepackage{apjfonts}
\usepackage{url}
\usepackage[utf8]{inputenc}
\usepackage{fancyref}
\usepackage{hyperref}
\hypersetup{colorlinks=false,pdfborder={0 0 0}}

\shorttitle{Polarization measurements of hot dust stars}
\shortauthors{J.~P. Marshall et al.}

\newcommand{\sdssg}{$g^{\prime}$}
\newcommand{\sdssr}{$r^{\prime}$}
\newcommand{\icarus}{$Icarus$}

\begin{document}

\title{Polarization measurements of hot dust stars\\ and the local interstellar medium}

\author{J. P. Marshall\altaffilmark{1,2}}
\author{D. V. Cotton\altaffilmark{1,2}}
\author{K. Bott\altaffilmark{1,2}}
\author{S. Ertel\altaffilmark{3,4}}
\author{G. M. Kennedy\altaffilmark{5}}
\author{M. C. Wyatt\altaffilmark{5}}
\author{C. del Burgo\altaffilmark{6}}
\author{O. Absil\altaffilmark{7}}
\author{J. Bailey\altaffilmark{1,2}}
\author{L. Kedziora-Chudczer\altaffilmark{1,2}}

\altaffiltext{1}{School of Physics, UNSW Australia, High Street, Kensington, NSW 2052, Australia}
\altaffiltext{2}{Australian Centre for Astrobiology, UNSW Australia, High Street, Kensington, NSW 2052, Australia}
\altaffiltext{3}{Steward Observatory, Department of Astronomy, University of Arizona, 933 N. Cherry Ave, Tucson, AZ 85721, USA}
\altaffiltext{4}{European Southern Observatory, Alonso de Cordova 3107, Vitacura, Casilla 19001 Santiago, Chile}
\altaffiltext{5}{Institute of Astronomy, University of Cambridge, Madingley Road, Cambridge, CB3 0HA, UK}
\altaffiltext{6}{Instituto Nacional de Astrof\'isica, \'Optica y Electr\'onica, Luis Enrique Erro 1, Sta. Ma. Tonantzintla, Puebla, Mexico}
\altaffiltext{7}{Institut d'Astrophysique et de G\'eophysique, University of Li\`ege, 19c all\'ee du Six Ao\^ut, 4000 Li\`ege, Belgium}

\begin{abstract}
Debris discs are typically revealed through excess emission at infrared wavelengths. Most discs exhibit excess at mid- and far-infrared wavelengths, analogous to the solar system’s Asteroid and Edgeworth-Kuiper belts. Recently, stars with strong ($\sim$ 1 per cent) excess at near-infrared wavelengths were identified through interferometric measurements. Using the HIgh Precision Polarimetric Instrument (HIPPI), we examined a sub-sample of these hot dust stars (and appropriate controls) at parts-per-million sensitivity in SDSS g$^{\prime}$ (green) and r$^{\prime}$ (red) filters for evidence of scattered light. No detection of strongly polarized emission from the hot dust stars is seen. We therefore rule out scattered light from a normal debris disk as the origin of this emission. A wavelength-dependent contribution from multiple dust components for hot dust stars is inferred from the dispersion (difference in polarization angle in red and green) of southern stars. Contributions of 17 ppm (green) and 30 ppm (red) are calculated, with strict 3-$\sigma$ upper limits of 76 and 68 ppm, respectively. This suggests weak hot dust excesses consistent with thermal emission, although we cannot rule out contrived scenarios, e.g. dust in a spherical shell or face-on discs. We also report on the nature of the local interstellar medium, obtained as a byproduct of the control measurements. Highlights include the first measurements of the polarimetric colour of the local interstellar medium and discovery of a southern sky region with a polarization per distance thrice the previous maximum. The data suggest $\lambda_{\rm max}$, the wavelength of maximum polarization, is bluer than typical.
\end{abstract}


\section{Introduction}
\label{sec:intro}

The presence of remnant material from planet formation processes is most commonly revealed around mature, main sequence stars by detection of excess emission, above the stellar photospheric emission, at infrared wavelengths \citep{Wyatt2008,Matthews2014}. The origin of the observed excess is micron- to millimetre-sized dust grains produced in collisions between larger, unseen bodies \citep{BacPar1993,Krivov2010}. Hence these objects are called debris discs. Typically, observed dust emission peaks at either mid- or far-infrared wavelengths with temperatures, derived from blackbody fits to the excess, of $\sim$200~K or 80~K \citep{Morales2011}; such discs are analogous to the Asteroid or Edgeworth-Kuiper belts, respectively. Hundreds of stars have been identified as debris disk host stars, many of which require multiple debris components echoing the structure of our solar system \citep{Chen2009,Su2013,KenWya2014}. For cool debris discs, i.e. Edgeworth-Kuiper belt analogues, that are more easily detected in contrast to their host stars, an incidence of $\sim$~20 to 30~per cent for AFGK-type stars has been measured \citep{Eiroa2013,Thureau2014}, albeit with limitations due to instrument sensitivity and survey strategy.

Interferometric measurements of Vega, the archetype of debris disk systems \citep{Aumann1984}, measured a visibility deficit in the near-infrared $H$ and $K$ wavebands \citep{Absil2006,Defrere2011}. This deficit was interpreted as being the result of emission originating within the field of view ($\sim$~200~mas), but at spatial scales of several stellar radii, i.e. more extended than the compact stellar emission. The favoured explanation for this emission was the presence of hot dust, with temperatures in excess of 1000~K, around the host star. Subsequent studies of other systems spanning a broad range of spectral types found similar excesses, some curiously around stars without any other evidence of excess emission \citep{diFolco2007,Absil2008,Absil2009,Akeson2009,Defrere2012}. Surveys combining measurements in $H$ and $K$ bands found an incidence of hot dust comparable to that of cooler debris discs around main sequence stars, but uncorrelated with the presence of cooler debris \citep{Absil2013,Ertel2014}.

The interpretation of the excess as being caused by thermal emission from dust is problematic. Any such grains would be close to the sublimation temperature, at small separations of $\sim$~0.2~au. The presence of grains with properties commonly ascribed to those in cooler belts is also problematic due to the short lifetime against collisional destruction, or removal from the system by radiation pressure \citep{Burns1979,Krivov2010}. Dust grains with temperatures ascribed to hot dust would also emit strongly in the mid-infrared, which is not seen \citep{Mennesson2014}, nor supported by modelling \citep{vanLieshout2014}. Delivery of sufficient dusty material to the star's vicinity from exterior debris belts, e.g. by high orbital eccentricity comets, is likewise difficult to achieve from a dynamical perspective \citep{Bonsor2012,Bonsor2013}. However, some scenarios have been proposed that account for the flaws noted above through e.g. the adoption of more realistic sublimation physics for dust grains \citep{Lebreton2013}, or the trapping of small, nano-scale dust grains in the stellar magnetic field \citep{Su2013,Rieke2015}. Regardless of its origin, at least in the cases of HD~20794 and HD~38858, the dust responsible must be located close to the host star \citep{Kennedy2015b}. In the interest of completeness we note that the hot dust phenomenon as a byproduct of the host star's properties has been considered, e.g. winds \citep{Absil2008} or oblateness \citep{Akeson2009}; such mechanisms have generally been ruled out as the cause of this phenomenon. However, it is clear that alternative mechanisms to account for the presence of these excesses should be explored. Here we investigate the possibility that the near-infrared excesses are due to scattered light, not thermal emission, from dust, while at the same time setting further observational constraints on the grains if the excesses are thermal emission.

A further motivation to examine a scattered light origin of the observed excess is the direct imaging of habitable zone terrestrial exoplanets \citep{Agol2007,Beckwith2008}. In the Solar system, debris dust migrating from the Asteroid Belt and deposited by comets is pervasive \citep[e.g.][]{Dermott1984,Nesvorny2010}. Sunlight scattered by these dust grains produces the Zodiacal light, the faint ($L_{\rm dust}/L_{\star}~\sim~10^{-7}$) inner component of the Solar system's debris disk \citep{BacPar1993}. Around other stars, the presence of a strong scattered light background from dust at separations of a few au from the host star constitutes a bright background from which the light from an exoplanet must be disentangled \citep{Roberge2012,Stark2015}. Ground-based efforts to characterize this emission for Sun-like stars with no infrared excess determined that most (95 per cent) of these exo-zodis are less than 60 times brighter than that of the Solar system in the mid-infrared \citep{Mennesson2014}. The Large Binocular Telescope Interferometer will be sensitive in the mid-infrared to exo-zodis only a few times brighter than the Solar system around nearby Sun-like stars, probing direct analogues of the Solar system \citep{Weinberger2015,Kennedy2015a}. Deriving constraints on the optical scattered light brightness from these limits is non-trivial. This is critical, however, since direct imaging searches for exo-Earths are supposed to be carried out at such wavelengths. 

The detected near-infrared excesses are potentially problematic for the direct detection of exo-Earths, since these might be explained by strong scattered-light emission \citep{Ertel2014}. To test this hypothesis we have measured the degree of polarization of six hot dust stars without any notable excess at mid- or far-infrared wavelengths using the HIgh Precision Polarimetric Instrument, a parts-per-million (ppm, $\times10^{-6}$) sensitivity aperture polarimeter \citep[HIPPI;][]{Bailey2015}. A 1 per cent scattered light excess in the near-infrared from micron-sized dust grains should be detectable in polarization at visible wavelengths, assuming a conversion of between 5 to 50~per cent for the magnitude of polarization from scattered light brightness \citep[][Marshall in prep.]{Schneider2014}. As a lower limit, a 1 per cent conversion between near-infrared and optical scattered light brightness would produce a signal of 100~ppm. This is well within the measurement capabilities of HIPPI;  its 1-$\sigma$ sensitivity of 10~ppm for a 1 hour integration on a 6th magnitude star would obtain a detection of the hot dust at the 10-$\sigma$ level.

In Section \ref{sec:obs} we present our polarimetric observations. We present the results of our study in Section \ref{sec:res}, followed by a discussion of these results in relation to the current understanding of hot dust in Section \ref{sec:dis}. Finally, in Section \ref{sec:con}, we summarise our findings and present our conclusions. 

\section{Observations}
\label{sec:obs}

\begin{deluxetable*}{llcccrrrrrr}
\tablecolumns{11}
\tablecaption{Stellar properties, fractional near-infared excess (in per cent), and summary of HIPPI observations. \label{table:obs_log}}
\tablehead{
\colhead{Name} & \colhead{Excess/} & \colhead{Right Ascension} & \colhead{Declination} & $V$ & \colhead{Spectral}  & \colhead{$d$} & \colhead{\sdssg} & \colhead{} & \colhead{\sdssr} & \colhead{} \\
\colhead{}     & \colhead{Control}     & \colhead{[hh mm ss]} & \colhead{[dd mm ss]} & [mag] & \colhead{Type}          & \colhead{[pc]}     & \colhead{Date}            & \colhead{Time [s]}       & \colhead{Date}     & \colhead{Time [s]}
}
\startdata
HD~2262   & 0.67~$\pm$~0.18 & 00 26 12.2 & -43 40 47 & 3.94 & A5~IV   &  23.8  & 18/10/15 & 1280 & 31/10/15 & 1280 \\
HD~739    & Control         & 00 11 44.0 & -35 07 59 & 5.20 &  F5~V   &  21.3  & 18/10/15 & 1280 & 2x31/10/15 & 1600 \\
\hline
HD~28355  & 0.88~$\pm$~0.09  & 04 28 50.2 &  13 02 51 & 5.01 &  A7~V   &  48.9  & 18/10/15 & 1280 & 31/10,2/11/15 & 1920 \\
HD~28556  & Control          & 04 30 37.4 &  13 43 28 & 5.40 &  F0~V   &  45.0  & 2/11/15 & 1280 & 2/11/15 & 1280 \\
\hline
HD~187642 & 3.07~$\pm$~0.24  & 19 50 47.0 &  08 52 06 & 0.76 &  A8~V   &   5.1  & 18/10/15 & 640 & 1/11/15 & 640 \\
HD~187691 & Control          & 19 51 01.6 &  10 24 57 & 5.10 &  F8~V   &  19.2  & 29/10/15 & 960 & 2/11/15 & 960 \\
\hline
HD~7788   & 1.43~$\pm$~0.17  & 01 15 46.2 & -68 52 33 & 4.25 &  F6~V   &  21.0  & 19/10/15 & 800 & 29,31/10/15 & 1600 \\
HD~4308   & Control          & 00 44 39.3 & -65 38 58 & 6.60 &  G8~V   &  22.0  & 29/10/15 & 1280 & 31/10,2/11/15 & 2240 \\
HD~7693   & Control          & 01 15 01.0 & -68 49 08 & 7.24 &  K2~V   &  21.7  & 2/11/15 & 1280 & 2/11/15 & 1280 \\
\hline
HD~14412  & 0.96~$\pm$~0.21  & 02 18 58.5 & -25 56 44 & 6.34 &  G8~V   &  12.7  & 20/10/15 & 960 & 31/10,2/11/15 & 2240 \\
HD~12311  & Control          & 01 58 46.2 & -61 34 12 & 2.84 & F0~IV   &  22.0  & 2/\phantom{0}9/14 & 640 & 31/10/15 & 640 \\
\hline
HD~210302 & 0.83~$\pm$~0.25  & 22 10 08.8 & -32 32 54 & 4.92 &  F6~V   &  18.3  & 29/10/15 & 960 & 2/11/15 & 960 \\
HD~176687 & Control          & 19 02 36.7 & -29 52 48 & 2.61 &  A2~V   &  27.0  & 1/\phantom{0}9/14 & 640 & 2/11/15 & 640 \\
\enddata
\vspace{-0.3cm}
\end{deluxetable*}

We observed six hot dust stars and seven control stars in the SDSS \sdssg~and \sdssr~filter bands with the HIgh Precision Polarimetric Instrument \citep[HIPPI; ][]{Bailey2015} on the 3.9-m Anglo-Australian Telescope (AAT). The hot dust stars examined here have no evidence of excess emission at mid- or far-infrared wavelengths, with limits on the fractional excess of dust ($L_{\rm dust}/L_{\star}$) of 10$^{-5}$ to 10$^{-6}$. The scattered light emission from any cold dust around these stars must therefore be negligible due to the limits from the fractional luminosity. The presence of any polarization from a star can thus be attributed solely to a combination of the interstellar medium (ISM) and the presence of circumstellar hot dust. We cannot rule out the presence of hot dust for any of the control stars, but the incidence of this phenomenon is $\sim$~10 per cent and from binomial probability we therefore expect at most one of the seven control stars (37 per cent probability) is actually a hot dust star.

A summary of the observations, including the new observations obtained for this programme in two observing runs during 2015 (14/10-20/10 and 29/10-3/11), is given in Table \ref{table:obs_log}. The integration time per target for each Stokes parameter, $U$ and $Q$, is half the total time listed in the table. The two observations made in September of 2014 were reported previously in \citet{Cotton2016}, but the details are included here for completeness.

HIPPI is a high precision polarimeter, with a reported sensitivity in fractional polarization of $\sim~4.3~$ppm on stars of low polarization and a precision of better than 0.01~per cent on highly polarized stars \citep{Bailey2015}. It achieves this by the use of Ferroelectric Liquid Crystal (FLC) modulators operating at a frequency of 500~Hz to eliminate the effects of variability in the atmosphere. Second stage chopping, to reduce systematic effects, is accomplished by rotating the entire back half of the instrument after the filter wheel, with a typical frequency of once per 20 seconds.

Observations in the \sdssg~filter used the blue sensitive Hamamatsu H10720-210 Ultra bialkali photocathode photomultiplier tube (PMT), as per previously reported observations with HIPPI \citep{Bailey2015,Cotton2016}. Observations made with the \sdssr~filter used the red sensitive Hamamatsu H10720-20 infrared extended multialkali photocathode to improve the efficiency of measurements made in \sdssr. Using the bandpass model described in \citet{Bailey2015} we have determined the effective wavelength and efficiency for various spectral types without reddening for this filter-PMT combination, and this is presented in Table \ref{table:filter}. We also reproduce the same data for the \sdssg~filter and Ultra-bialkali PMT combination for completeness.

\begin{deluxetable}{lcccc}
\tablecolumns{5}
\tablecaption{Effective wavelength and modulation efficiency for different spectral types according to our bandpass model. \label{table:filter}}
\centering
\tablehead{
\colhead{Spectral} & \colhead{\sdssg} & \colhead{} & \colhead{\sdssr} & \colhead{} \\
\colhead{Type}     & \colhead{$\lambda_{\rm eff}$ [nm]} & \colhead{Mod. Eff.} & \colhead{$\lambda_{\rm eff}$ [nm]} & \colhead{Mod. Eff.}}
\startdata
B0 & 459.1 & 0.877 & 616.8 & 0.814 \\
A0 & 462.2 & 0.886 & 618.3 & 0.811 \\
F0 & 466.2 & 0.896 & 620.8 & 0.807 \\
G0 & 470.7 & 0.906 & 623.0 & 0.802 \\
K0 & 474.4 & 0.916 & 624.5 & 0.800 \\
M0 & 477.5 & 0.920 & 629.3 & 0.791 \\
M5 & 477.3 & 0.917 & 630.4 & 0.789 \\
\enddata
\vspace{-0.3cm}
\end{deluxetable}

A sky measurement, lasting 40 seconds, was acquired at each of the four telescope position angles an object was observed at, and subtracted from the measurement. The sole exception was HD~12311 in the \sdssg~filter for which, being a particularly bright object observed in good conditions, a dark measurement was sufficient for calibration purposes. These subtractions were carried out as the first part of the data reduction routine, that determines polarization via a Mueller Matrix method. Full details are provided by \citet{Bailey2015}.

Angular calibration was carried out with reference to a set of high polarization standards with known polarization angles: HD~23512, HD~187929, HD~154445 and HD~80558. The standards have angles known to a precision of $\sim$1$\degr$ -- which dominates the uncertainty of our measurements. During the observations, zero point calibration (telescope polarization; hereafter abbreviated TP) was carried out by reference to the average of a set of observed stars with measured low polarizations; this is shown in Table \ref{table:tp}. Note that the \sdssg~HIP~2021 observation of the 29th of October and the \sdssg~Sirius observations of 2nd November were not used as part of the calibration for \sdssg~observations carried out during the 14-20 October run. The reason being that the telescope polarization can drift over time, and so where sufficient measurements are available, contemporaneous calibration observations are preferred. Owing to challenging conditions only two additional calibration observations were possible for \sdssg~in the later run, and so all the measurements from both runs were used. The difference in the calibration is within the reported error, and we consider it of no consequence.

\begin{deluxetable}{llcc}
\tablecolumns{4}
\tablecaption{Telescope polarization (TP) measurements for the October and November 2015 runs in \sdssg~and \sdssr~filters. \label{table:tp}}
\centering
\tablehead{
\colhead{Star} &  \colhead{Date} &     \colhead{$p$ [ppm]} &   \colhead{\hspace{2 mm}$\theta$ [$\degr$]}}
\startdata
HIP~2021 &    14 Oct &   58.1 $\pm$ 4.1 & 90.5 $\pm$ 4.0    \\
         &    19 Oct &   52.5 $\pm$ 3.9 & 91.4 $\pm$ 4.3    \\
         &    29 Oct &   56.0 $\pm$ 4.1 & 88.2 $\pm$ 4.2    \\
Sirius   &    16 Oct &   52.8 $\pm$ 0.7 & 88.2 $\pm$ 0.8    \\
         &    19 Oct &   49.8 $\pm$ 1.3 & 90.7 $\pm$ 1.8    \\
         &     2 Nov &   47.3 $\pm$ 1.1 & 86.5 $\pm$ 0.6    \\
\hline
Adopted TP & \sdssg & 55.9 $\pm$ 1.1 & 89.3 $\pm$ 0.6    \\
\hline
HIP~2021 &   31 Oct  &   41.0 $\pm$ 5.2 & 93.0 $\pm$ 3.6    \\ 
Sirius   &   29 Oct  &   31.8 $\pm$ 6.2 & 96.3 $\pm$ 6.4    \\ 
         &    2 Nov  &   31.8 $\pm$ 1.5 & 93.6 $\pm$ 1.3    \\
\hline
Adopted TP & \sdssr & 34.8 $\pm$ 2.5 & 94.2 $\pm$ 2.2   \\
\enddata
\vspace{-0.3cm}
\end{deluxetable} 

We set out to observe a set of six stars for the ISM control sample, also with no known excess emission, with spectral types unlikely to be intrinsically polarized, that were of similar brightness to the target stars and reasonably closely placed on the sky (within a few degrees). All of the control stars would therefore have comparable precision in their polarimetry measurements to the hot dust stars. No effort was made to match the spectral type of the target-control pairs, which leads to slight mis-matches between the effective wavelength of measurement for polarization between hot dust and control stars. These differences are all small ($<$~10~nm), and can therefore be neglected. The relative distances of the target-control pairs were not able to be precisely matched. To compensate for this we assumed that the polarization induced by the interstellar medium was linearly related to distance and scaled the polarizations measured for control stars by the ratio of the target and control distances for our analysis. 

In one case, where the control was particularly polarimetrically red and the correpsonding target star recorded a particularly high degree of polarization for its distance from the Sun, we observed a second nearby star to confirm the polarization was interstellar in origin –- later we treat the distance-scaled average of these two stars as a single control. In two further cases, owing to particularly challenging observing conditions, we had to substitute brighter stars for controls. These two stars are not as close to the corresponding target as the rest of the set. However, in this instance they are both the nearest stars that had been observed previously with HIPPI and shown to be polarized only by the ISM, and they both had a similar polarization per distance to the targets \citep{Cotton2016}. 

\section{Methodology and results}
\label{sec:res}

\subsection{Methodology}

For the analysis of the data, we split the stars according to their nature (i.e. target or control), and further subdivide them according to the target star's spectral type (i.e. A stars and F/G stars). The grouping of stars into A and F/G sub-groups is done to facilitate a comparison of any measureable polarization properties as a function of spectral type, where dust grain size may become a factor. To compare these groupings we calculate the ratio of their polarization in the two filter bands, and the consistency of the orientation angles in each filter band. We have examined the stars in aggregate rather than individually, because the signal-to-noise of polarization measurements is generally low and the clumpiness of the ISM makes vector subtraction of polarization contributions unreliable as a mechanism to assess the relative contribution of components toward individual stars. In Table \ref{tab:pol_results} we summarise the individual measurements for the stars observed in this work.

Our method for determining the error in normalised Stokes parameters, $q$ ($Q/I$) and $u$ ($U/I$), with HIPPI is given in detail in \cite{Bailey2015}. The error in $p$, $\sigma_{p}$, is simply the mean of $\sigma_{\rm q}$ and $\sigma_{\rm u}$. These errors decrease as $\sqrt{n}$ for the mean of $n$ individual measurements. Where we have made multiple observations of an object, the means and errors of $q$ and $u$ have been calculated by weighting of the errors in the individual observations. In particular this applies to the raw data presented in Table \ref{tab:pol_results}.

Whilst the error in the magnitude of polarization is straightforward to calculate, the error in polarization angle, $\sigma_{\theta}$, requires more care. If the signal to noise ratio $p/\sigma_{p}$ is large then the probability distribution function for $\theta$ is Gaussian, and 1$\sigma$ errors (in degrees) are given by \citet{Serkowski1962}: \begin{equation} \label{eq_S_PA_Gau_Err} \sigma_{\theta} = 28.65~\sigma_{p}/p, \end{equation}  However when $p$/$\sigma_{p}~<~$4 the distribution of $\theta$ becomes kurtose with appreciable wings. In such cases Equation \ref{eq_S_PA_Gau_Err} is no longer strictly accurate -- though it is often still used when precision is not critical (the difference is generally much less than 5 degrees). Throughout this work we have taken extra care and made use of the work of \citet{Na-KhCla1993} who give precisely $\sigma_{\theta}$ as a function of $p$/$\sigma_{p}$ in their Figure 2(a).

In contrast to the normalised Stokes parameters $q$ and $u$, the magnitude of polarization, $p$, is positive definite; this needs to be considered when taking the mean of many values of $\bar{p}$, especially if $\bar{p}$/$\bar{\sigma_{p}}$ is low, as in Table \ref{tab:sample_results}. The standard method for debiasing such data, to best estimate the true value of, $p$, is that of \citet{Serkowski1962}: \begin{equation} \label{eq_S_debias} \hat{p} \sim (\bar{p}^2-\bar{\sigma_{p}}^2)^{1/2},\end{equation} where $\hat{p}$ is the debiased polarization, and $\bar{\sigma_{p}}$ is calculated as root mean square error. \citet{WarKron1974} derive this same equation in a different way, and recommend its use for $\bar{p}$/$\bar{\sigma_{p}}$ $>$ 0.5. We apply this debiasing when we consider trends in $p$ in an ensemble of data. 

The observables we have drawn for our sample are the magnitude and angle of polarization. In Table \ref{tab:sample_results} we aggregate the individual measurements and compare the hot dust stars to the controls both in total and by spectral type (distinguishing A and F/G type stars). To reduce the effects of distance variation and better compare the controls to the targets, we scale the polarization measurements of each of the controls linearly by the target:control distance ratio (implicitly assuming that the ISM polarization magnitude varies linearly in distance).

To analyse the influence of hot dust on the observed polarization we:
\begin{itemize}
\item compare the magnitude of polarization for the hot dust and control stars in each of the \sdssg~(green) and \sdssr~(red) bands,
\item calculate the ratio of polarization magnitudes $p_{\rm green}$:$p_{\rm red}$,
\item calculate the dispersion ($\phi$) -- the difference in orientation angle between green and red ($\theta_{\rm green}$ and $\theta_{\rm red}$ -- and compare the hot dust and control star groups.
\end{itemize}

\subsection{Results}

\subsubsection{Non-detection of strongly polarized light}
\label{sssec:no_pol}

Our primary finding is the non-detection of strong polarized light from the hot dust stars. The magnitude of polarization for the hot dust stars is found to be consistent with that of the distance-scaled control stars. To gain an idea of how large the polarizing effect of hot dust is likely to be if present, we can make a simple calculation based on the statistics of vector addition. If we take the average distance-scaled polarization of the controls to be representative of the interstellar polarization, $\bar{p_{\rm i}}$, and the hot dust targets to be $p = p_{\rm i} + p_{\star}$, where $p_{\star}$ is the intrinsic polarization of the system, then for a large enough sample the median intrinsic polarization is given by: \begin{equation}\label{eq_median_p} \hat{p_{\star}} = \sqrt{\bar{p}^2-\bar{p_{i}}^2}. \end{equation} Using the sample of southern hot dust stars and controls (since the dispersion measurements lead us to believe the ISM in the south is less patchy -- see below), this equation produces 17~$\pm$~20~ppm in green, and 30~$\pm$~13~ppm in red\footnote{These numbers are consistent with the hypothesis that a hot dust component in HD~28355 is anti-aligned with an ISM component to produce its very low polarization.}. From this we obtain 3-$\sigma$ upper limits of 76~ppm in green and 68~ppm in red. These are statistical limits, but if the hot dust contribution had been at the 100~ppm level for any star it would have dominated the ISM component. We therefore infer that the contribution from the hot dust and ISM are of approximately the same magnitude. If the hot dust contribution had been the dominant component, its presence would have been much easier to detect than the derived upper limits suggest.

If the near-infrared excess is assumed to be scattered light, and this scattering is wavelength independent, then the expected polarization is simply the fractional excess $F_{\rm IR}/F_{\star}$ multiplied by the polarization fraction of light scattered by the dust $f_{\rm pol}$: \begin{equation}\label{eq_scat_est} p = (F_{\rm IR}/F_{\star})f_{\rm pol} \end{equation} The fractional excesses of the hot dust stars lie at the $\sim$1 per cent level \citep{Absil2013,Ertel2014}. For consistency with the upper limits obtained from HIPPI measurements (78~ppm in green, 68~ppm in red), $f_{\rm pol}$ must lie at the $\leq~$1 percent level, i.e. the dust grains must be very strongly non-polarizing. For debris disks resolved in scattered light \citep{Schneider2014}, the polarization fraction of the scattered light lies between 5 and 50 per cent (Marshall et al. in prep.), leading to expected polarization signals of $\geq$~500~ppm from these hot dust stars; such signals would have been clearly detected in the data obtained here. If the dust grains were more red in their scattering colour the excess at optical wavelengths would be reduced from that measured at near-infrared wavelengths. 

\subsubsection{Polarimetric colour of the local ISM}

The wavelength dependence of polarization of the ISM usually peaks at optical wavelengths \citep{Serkowski1975}. The empirical wavelength dependence of interstellar polarization is given by the Serkowski Law \citep{Serkowski1975} as modified by \citet{Wilking1982}: \begin{equation} \label{eq_serkowski} p(\lambda)/p(\lambda_{max})=\exp((0.1-1.86\lambda_{max})\ln^{2}(\lambda/\lambda_{max})),\end{equation} where $\lambda$ is the wavelength examined and $\lambda_{max}$ the wavelength of maximum polarization.

\begin{figure}
\centering
\includegraphics[width=0.5\textwidth]{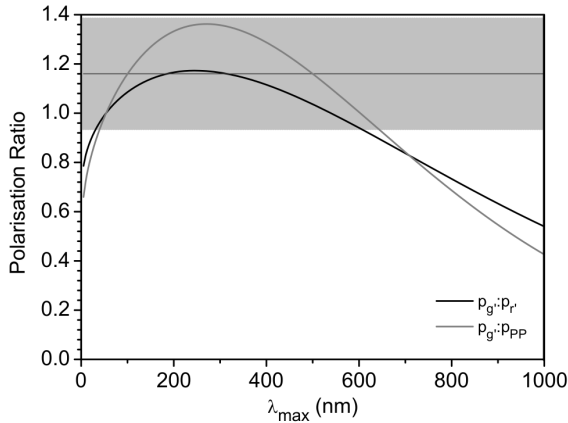}
\caption{Plot of the $p_{\rm green}$:$p_{\rm red}$ ratio (black) vs. $\lambda_{max}$ determined from Serkowski's Law \citep{Serkowski1975,Wilking1982}. The shaded region corresponds to the 1-$\sigma$ error in the mean $p_{\rm green}$:$p_{\rm red}$ ratio obtained from the control stars. To show the conversion implied for PlanetPol data, relevant for Fig. \ref{fig:sky_plot}, the $p_{\rm green}$:$p_{\rm PlanetPol}$ ratio vs. $\lambda_{max}$ is also shown (grey). \label{fig:ratio_vs_lambdamax}}
\end{figure}

Using the green and red filter measurements of the southern control stars we determined a mean ISM polarimetric colour and fit the Serkowski Law to it. We have restricted this analysis to southern stars because the increased interstellar polarization at southern latitudes reduces the influence of statistical uncertainties. Additionally, as discussed later, the southern controls, as a group, display minimal dispersion. The mean distance of these stars is 27.4~pc. The mean effective wavelength of the controls in the \sdssg and \sdssr~filters are 468.0~nm and 621.6~nm, respectively. The Serkowski Law allows us to determine the ratio of interstellar polarization at these two wavelengths for any given $\lambda_{max}$, simply by calculating Equation \ref{eq_serkowski} for each and taking the ratio, i.e. \begin{equation}p_{\rm green}:p_{\rm red}=\frac{(p(468.0~nm)/p(\lambda_{max})}{(p(621.6~nm)/p(\lambda_{max})} \end{equation} The maximum possible $p_{\rm green}$:$p_{\rm red}$ ratio may be obtained simply by plotting $p_{\rm green}$:$p_{\rm red}$ against $\lambda_{max}$ as in Fig. \ref{fig:ratio_vs_lambdamax}. In this case the maximum ratio is 1.17, and occurs for a $\lambda_{max}$ of 250~nm. The ratio we measured from the controls was close to this: 1.16~$\pm$~0.22. As can be seen from Fig. \ref{fig:ratio_vs_lambdamax} this corresponds to $\lambda_{max}$ being equal to either $\sim$~185~nm or $\sim$~315~nm, or considering the uncertainty, a range of possible values for $\lambda_{max}$ between $\sim$~35~nm and $\sim$~600~nm. 

\subsubsection{Hot dust}

\begin{figure*}
\centering
\includegraphics[width=\textwidth,trim={0 5cm 0 1cm}]{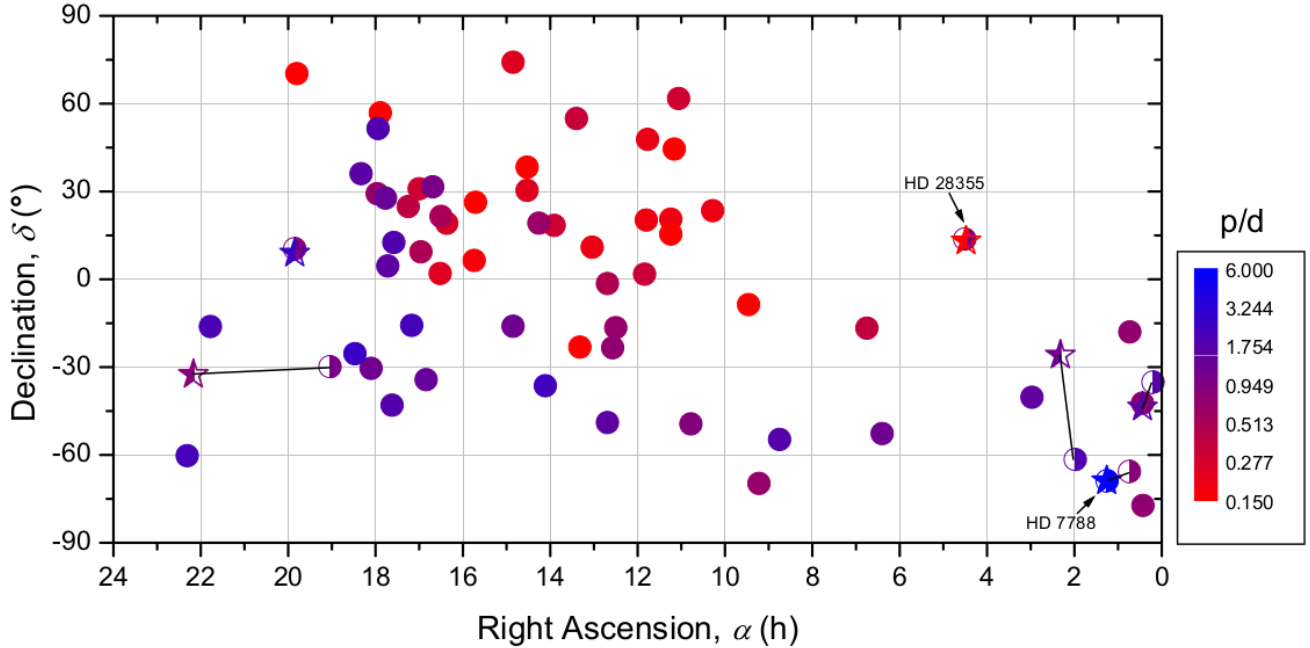}
\caption{Plot of polarization/distance ($p/d$) vs. sky position for the hot dust and control stars in the \sdssg~filter. Target stars are half-filled star shapes, whilst controls are half-filled circles. Target-control pairs are connected by solid lines. Literature measurements, shown as filled circles, are taken from \cite{Cotton2016} and \cite{Bailey2010}. Only those stars believed to have negligible intrinsic polarization have been included. All data has been debiased according to Equation \ref{eq_S_debias}. The PlanetPol values have been scaled to \sdssg~according to the mean colour of the ISM determined from our \sdssg~and \sdssr~measurements using Serkowski's Law; see text for details. Although the colour scale runs from p/d of 0.15 to 6.0, some stars plotted fall outside this range, and are shown as the extreme colour. Note that although some literature stars appear closer to the hot dust stars than the selected controls when projected on the sky, they are further away when taking account of distance. \label{fig:sky_plot}}
\end{figure*}

In Fig. \ref{fig:sky_plot} we show the magnitude of polarization in green for all stars scaled by the relative stellar distances (i.e. $p/d$), as compared to stars from the HIPPI \citep{Cotton2016} and PlanetPol \citep{Bailey2010} surveys. The PlanetPol stars were observed at a longer wavelength than those observed by HIPPI (roughly 750~nm cf 460~nm for an F0 star), and thus the PlanetPol measurements must be scaled accordingly to compare the two surveys. Since the value for $\lambda_{max}$ of 315~nm determined in the previous section results in a near maximal conversion ratio, and has a large uncertainty, we have taken account of the uncertainty rather than make a calculation based on 315~nm. Instead we calculated an expectation value of 1.20 for the $p_{\rm green}$:$p_{\rm PlanetPol}$ ratio based on normalised probabilities for a range of $\lambda_{max}$ values, where the probabilities result from an assumed normal distribution for the $p_{\rm green}$:$p_{\rm red}$ ratio with mean 1.16 and standard deviation 0.22. The calculated ratio corresponds to $\lambda_{max}$ equal to 470~nm -- a fairly conservative interpretation of the results.

Broadly speaking, the target stars have polarizations consistent with that expected from the ISM given their distances based on previous estimates. After scaling the magnitude of polarization of the control stars for the difference in distance between the control and its target counterpart their values do not differ greatly (see Table \ref{tab:pol_results}). The distance-scaled polarization ($p/d$) of the targets and controls as a function of their sky position is presented in Fig. \ref{fig:sky_plot}. Here we see that the distance-scaled polarization varies smoothly across the sky and that the polarization magnitude values obtained here for both targets and controls are consistent with what would be expected given the stars' sky positions and distances, as shown in \cite{Cotton2016}. In Fig. \ref{fig:sky_plot} it can be seen that there are sometimes closer stars already observed by HIPPI than those selected as controls in this work. Those stars are considered unsuitable for comparison with the hot dust stars due to the large disparity in distances between them. One exception to this is the case of HD~28355/HD~28556 in the green filter band, where (with large uncertainties) the control, HD~28556, exhibits polarization at the 50~ppm level, whereas the target, HD~28355, has no significant polarization. Given the distances to these targets ($d~\sim~50$~pc), we would expect an ISM contribution of $\sim~50~$ppm. It is possible that the absence of measurable polarization from HD~28355 is thus the result of multiple vector contributions cancelling each other out. Another exception is the large $p/d$ value seen for HD~7788 and HD~7693.

\begin{deluxetable*}{lllcrrrrrr}
\centering
\tablecolumns{10}
\tablecaption{Raw and distance modulated polarimetric measurements of hot dust and control stars. \label{tab:pol_results}}
\tablehead{
\colhead{Filter} & \colhead{HD} & \colhead{T/C} & \colhead{$\lambda_{\rm eff}$} & \colhead{$q$} & \colhead{$u$} & \colhead{$p$} & \colhead{$d$} & \colhead{$p/d$} & \colhead{$\theta$} \\
\colhead{} & \colhead{} & \colhead{} & \colhead{[nm]} & \colhead{[ppm]} & \colhead{[ppm]} & \colhead{[ppm]} & \colhead{[pc]}& \colhead{[ppm/pc]} & \colhead{[\degr]}
}
\startdata
\sdssg	&	2262	&	Target	&	464.2	&	  4.7~$\pm$~5.7	  &	37.4~$\pm$~5.7	&	37.7~$\pm$~5.7	& 23.8 & 1.58 &	41.4~$\pm$~4.3	\\
	    &	739	    &	Control	&	468.5	&	-31.0~$\pm$~12.6  &	24.3~$\pm$~12.4	&	39.3~$\pm$~12.5	& 21.3 & 1.84 &	71.0~$\pm$~9.6	\\
\sdssr	&	2262	&	Target	&	619.6	&	-7.3~$\pm$~8.4	  &	17.5~$\pm$~8.8	&	18.9~$\pm$~8.6	& 23.8 & 0.79 &	56.4~$\pm$~15.4	\\
	    &	739	    &	Control	&	621.9	&	-5.3~$\pm$~11.1	  &	24.3~$\pm$~11.3	&	24.9~$\pm$~11.2	& 21.3 & 1.17 &	51.1~$\pm$~15.3	\\
\hline
\sdssg	&	28355	&	Target	&	465.0	&	 0.0~$\pm$~10.7	&	-2.1~$\pm$~12.1	 &	2.1~$\pm$~11.4	& 48.9 & 0.04 &	134.6~$\pm$~48.2	\\
	    &	28556	&	Control	&	466.2	&	49.3~$\pm$~27.0	&	12.4~$\pm$~27.2	 &	50.8~$\pm$~27.1	& 45.0 & 1.13 &	 7.1~$\pm$~18.8	\\
\sdssr	&	28355	&	Target	&	620.1	&	-0.2~$\pm$~10.3	&	13.5~$\pm$~10.4	 &	13.5~$\pm$~10.4	& 48.9 & 0.28 &	45.5~$\pm$~26.7	\\
	    &	28556	&	Control	&	620.8	&	-3.9~$\pm$~13.9	&	-18.3~$\pm$~13.9 &	18.7~$\pm$~13.9	& 45.0 & 0.42 &	129.0~$\pm$~26.1	\\
\hline
\sdssg	&	187642	&	Target	&	465.4	&	13.3~$\pm$~2.7	 &	3.4~$\pm$~2.7	 &	13.7~$\pm$~2.7 & 5.1 & 2.69 &	7.2~$\pm$~5.8	\\
	    &	187691	&	Control	&	469.8	&	-9.9~$\pm$~10.7	 &	-16.1~$\pm$~10.6 &	18.9~$\pm$~10.6	& 19.2 & 0.98 &	119.3~$\pm$~19.9	\\
\sdssr	&	187642	&	Target	&	620.3	&	2.7~$\pm$~7.8	 &	-2.9~$\pm$~7.8	 &	4.0~$\pm$~7.8	& 5.1  & 0.78 &	156.5~$\pm$~41.6	\\
	    &	187691	&	Control	&	622.6	&	-10.4~$\pm$~17.4 &	20.1~$\pm$~17.6	 &	22.6~$\pm$~17.5	& 19.2 & 1.18 &	58.7~$\pm$~27.0	\\
\hline
\sdssg	&	7788	&	Target	&	468.9	&	-101.5~$\pm$~9.4	&	-39.1~$\pm$~9.6	  &	108.8~$\pm$~9.5   & 21.0 & 5.18 &	100.5~$\pm$~2.5	\\
	    &	4308	&	Control	&	473.7	&	-12.0~$\pm$~16.7	&	-21.4~$\pm$~16.3  & 24.6~$\pm$~16.5	  & 22.0 & 1.12 &	120.4~$\pm$~23.9	\\
	    &	7693	&	Control	&	475.0	&	-158.2~$\pm$~23.9	&	-38.3~$\pm$~23.6  &	162.8~$\pm$~23.7  & 21.7 & 7.50 &	96.8~$\pm$~4.5	\\
\sdssr	&	7788	&	Target	&	622.1	&	-113.0~$\pm$~11.6	&	-31.7~$\pm$~11.8  &	117.4~$\pm$~11.7  & 21.0 & 5.59 &	97.8~$\pm$~2.9	\\
	    &	4308	&	Control	&	624.2	&	-44.2~$\pm$~16.4	&	2.9~$\pm$~16.7	  &	44.3~$\pm$~16.5   & 22.0 & 2.01 &	88.1~$\pm$~11.8	\\
	    &	7693	&	Control	&	625.5	&	-147.0~$\pm$~25.2	&	-64.8~$\pm$~26.5  &	160.7~$\pm$~25.9  & 21.7 & 7.41 &	101.9~$\pm$~4.7	\\
\hline
\sdssg	&	14412	&	Target	&	473.7	&	-14.2~$\pm$~13.4  &	-14.4~$\pm$~12.6  &	20.2~$\pm$~13.0	& 12.7 & 1.59 &	112.7~$\pm$~22.9	\\
	    &	12311	&	Control	&	466.2	&	31.7~$\pm$~5.8	  &	-28.1~$\pm$~6.2	  &	42.4~$\pm$~6.0	& 22.0 & 1.93 &  159.2~$\pm$~4.1	\\
\sdssr	&	14412	&	Target	&	624.2	&	-37.4~$\pm$~17.1  &	27.2~$\pm$~17.1   &	46.2~$\pm$~17.1	& 12.7 & 3.64 &	72.0~$\pm$~11.7	\\
	    &	12311	&	Control	&	620.8	&	18.7~$\pm$~16.6   &	-27.6~$\pm$~17.8  &	33.4~$\pm$~17.2	& 22.0 & 1.52 &	152.0~$\pm$~18.1	\\
\hline
\sdssg	&	210302	&	Target	&	468.9	&	-15.5~$\pm$~12.6  &	-13.7~$\pm$~13.1  &	20.7~$\pm$~12.9	& 18.3 & 1.13 &	110.8~$\pm$~22.2	\\
	    &	176687	&	Control	&	463.2	&	0.9~$\pm$~4.6	  &	-28.2~$\pm$~4.7	  &	28.2~$\pm$~4.6	& 27.0 & 1.04 &	135.9~$\pm$~4.6	\\
\sdssr	&	210302	&	Target	&	622.1	&	8.7~$\pm$~12.9    &	6.4~$\pm$~14.3	  &	10.8~$\pm$~13.6	& 18.3 & 0.59 &	18.1~$\pm$~36.7	\\
	    &	176687	&	Control	&	618.9	&	-3.9~$\pm$~8.2    &	-10.0~$\pm$~8.3	  &	10.7~$\pm$~8.2	& 27.0 & 0.40 &	124.4~$\pm$~26.8	\\
\enddata
\vspace{-0.3cm}
\tablecomments{HD~187642 was also measured by PlanetPol \citep{Bailey2010}: $q = -7.3~\pm~1.3$, $u = -1.2~\pm~1.2$\\
For stars with two \sdssr measurements averaged above, we give the measurements of $q$ and $u$ for the individual observations as follows:\\
HD 739: $q = -11.5~\pm~16.7$, $u = 0.0~\pm~16.9$; $q = -0.4~\pm~14.9$, $u = 44.0~\pm~15.2$\\
HD 28355: $q = -26.7~\pm~18.2$, $u = 27.9~\pm~17.9$; $q = 12.4~\pm~12.6$, $u = 6.1~\pm~12.7$\\
HD 7788: $q = -104.3~\pm~16.9$, $u = -26.6~\pm~16.8$; $q = 120.8~\pm~16.0$, $u = -36.6~\pm~16.5$\\
HD 4308: $q = -47.5~\pm~23.7$, $u = -14.4~\pm~25.0$; $q = -41.2~\pm~22.6$, $u = 16.8~\pm~22.4$\\
HD 14412: $q = -55.2~\pm~27.0$, $u = 44.7~\pm~30.9$; $q = -25.3~\pm~22.2$, $u = 19.5~\pm~20.5$\\}
\end{deluxetable*}

For the A stars we find that the degree of green polarization is weaker in the hot dust stars than for the controls, at the 1.1-$\sigma$ level ($17.3~\pm~4.3$~ppm cf $32.9~\pm~10.9$~ppm, respectively). This may be an indication that the hot dust stars in this group are intrinsically polarized oppositely to the ISM, or that the ISM is particularly patchy. The polarization is consistent between the two groups in the red filter band ($11.0~\pm~5.2$~ppm cf $16.7~\pm~6.7$~ppm). Consequently the ratio of green to red is steeper for the control stars, 1.97~$\pm$~0.96, than for the A stars, 1.59~$\pm$~0.78, though due to the larger errors on the control group, not significantly so. Due to their proximity, and therefore weak polarization, the orientation angles have large uncertainties but both groups show tentative evidence for dispersion: $\phi = 44.9~\pm~23.7\degr$ for controls (1.9-$\sigma$), and $46.2~\pm~16.6\degr$ for targets (2.8-$\sigma$). Dispersion is indicative of multiple polarigenic mechanisms with different efficiencies at different wavelengths; its presence in the control group suggests multiple interstellar dust clouds with varying physical properties along the line of sight. Two of the three target-control pairs in this group are in the northern hemisphere where interstellar polarization is weaker and this scenario more likely. 

For the F/G stars we find that the red polarization of the hot dust stars is stronger than the controls, at the 1.4-$\sigma$ level ($57.5~\pm~8.3$~ppm cf $40.6~\pm~6.2$~ppm), but that the two groups are consistent in green ($49.4~\pm~6.9$~ppm cf $43.2~\pm~3.9$~ppm). The ratio of green to red for the targets, 0.86~$\pm$~0.17, is consistent with that of the controls, 1.06~$\pm$~0.20. All of these stars are situated in the southern hemisphere where the ISM contribution is stronger. Thus the contribution of hot dust to the measurement, if present, is further diluted. However, whilst the control group shows no dispersion ($\phi = 6.5~\pm~11.2\degr$), the F/G hot dust group does have a significant dispersion ($\phi = 43.6~\pm~16.5\degr$). This implies that there is an intrinsic component to the hot dust star polarization that has a different wavelength dependence to the ISM.

\begin{figure*}
\centering
\includegraphics[width=0.75\textwidth]{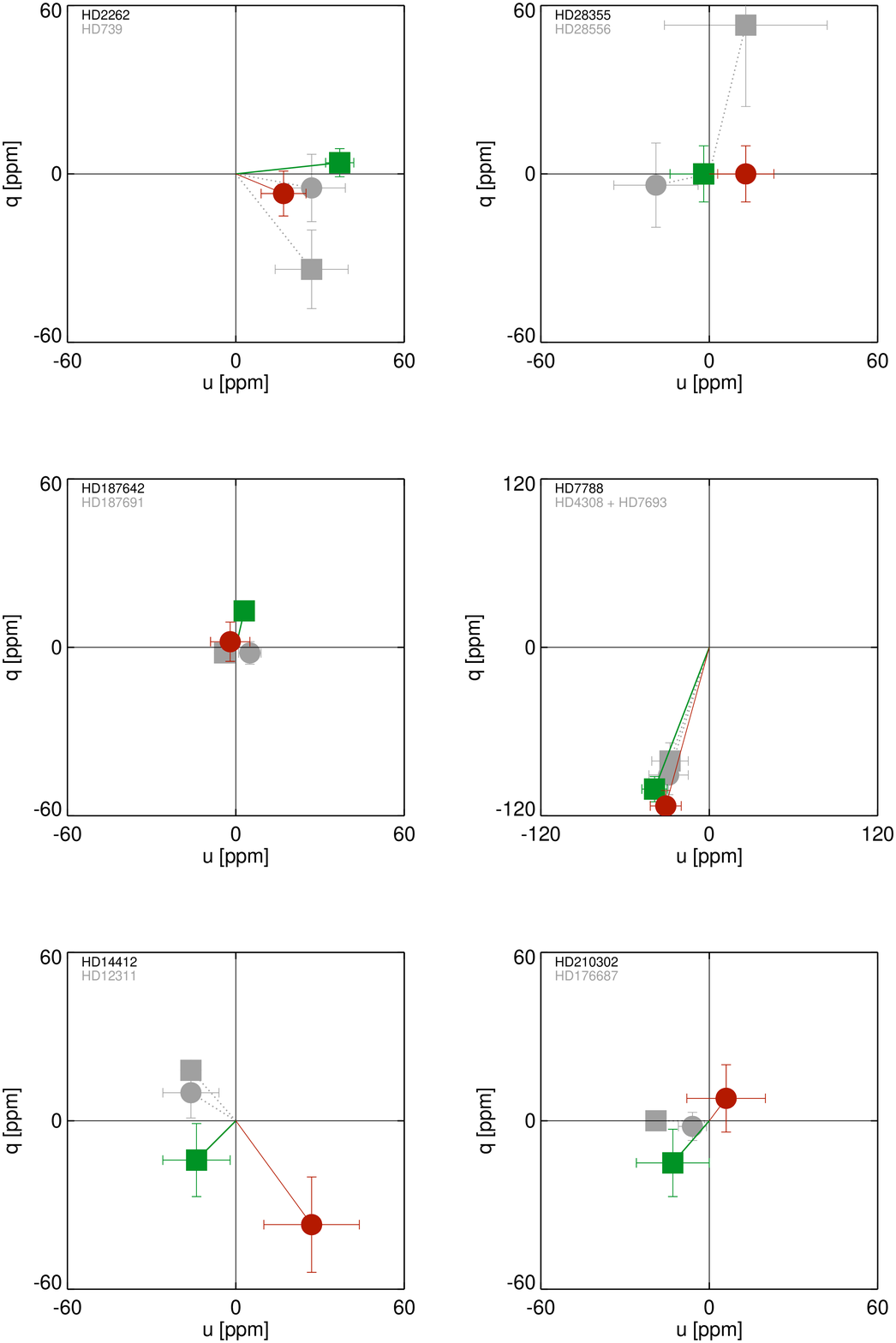}
\caption{Polarization plot showing $q$ and $u$ vectors to illustrate the dispersion of polarization between hot dust and control pairs in \sdssg~and \sdssr~filters. The controls have been scaled by their distance relative to the corresponding target. Axis scales are in parts-per-million. Squares denote \sdssg~measurements, whilst circles denote \sdssr~measurements. Target stars are shown in colour, whilst controls are in grey scale. Uncertainties are 1-$\sigma$. Solid lines denote the vector for target stars, whilst dashed lines denote the vector for control stars. The dispersion is half the angle between green and red filters as displayed on this plot. \label{fig:dispersion_plot}}
\end{figure*}

In aggregate, combining the A star and F/G star sub-samples, both the hot dust and control groups show dispersion ($\phi = 45.2~\pm~14.4\degr$ for targets and $37.0~\pm~10.0\degr$ for controls), but when the two northern control-target pairs are removed from the analysis, the remaining controls are consistent with only a very small level of dispersion ($\phi = 9.8~\pm~9.5\degr$, 1.0-$\sigma$) whilst the hot dust group still shows a more significant level of dispersion ($\phi = 36.4~\pm~13.0\degr$, 2.8-$\sigma$). The $p_{\rm green}$:$p_{\rm red}$ ratios suggest a stronger polarization in green for the ISM than for the hot dust stars (or, contra-wise, a rising contribution in red from hot dust stars compared to the controls), but this is not statistically significant. Thus the main result here is that there is no large significant polarization associated with the hot dust phenomenon. To illustrate the dispersion, the distance-scaled $q$ and $u$ vectors for each target-control pair are presented in Fig. \ref{fig:dispersion_plot}.

\begin{deluxetable}{llccc}
\tablecolumns{5}
\tablecaption{Comparison of hot dust and control samples. Control star measurements have been scaled by their distance relative to the corresponding control. Mean polarization measurements have been debiased according to Equation \ref{eq_S_debias}. \label{tab:sample_results}}
\tablehead{
\colhead{Filter} & \colhead{Sample} & \colhead{$\hat{p}$} & \colhead{$\phi$} & \colhead{Distance} \\
\colhead{} & \colhead{} & \colhead{[ppm]} & \colhead{[\degr]} & \colhead{[pc]}
}
\startdata
\sdssg	&	A	&	17.3~$\pm$~4.3	&	&			\\
	&	Control	&	32.9~$\pm$~10.9	&	&			\\
\sdssr	&	A	&	11.0~$\pm$~5.2	&	&			\\
	&	Control	&	16.7~$\pm$~6.7	&	&			\\
---	&	A Ratio	&	1.58~$\pm$~0.78	&	44.9~$\pm$~23.7	& 25.9\\
	&	Control Ratio	&	1.97~$\pm$~	0.96	&	46.2~$\pm$~16.6 &	\\
\hline
\sdssg	&	F/G	&	49.4~$\pm$~6.9	&	&			\\
	&	Control	&	43.1~$\pm$~5.0	&	&			\\
\sdssr	&	F/G	&	57.5~$\pm$~8.3	&	&			\\
	&	Control	&	40.6~$\pm$~6.2	&	&			\\
---	&	F/G Ratio	&	0.86~$\pm$~0.17	&	43.6~$\pm$~16.5	& 17.3\\
	&	Control Ratio	&	1.06~$\pm$~0.20	&	6.5~$\pm$~11.3	& \\
\hline
\sdssg	&	All	&	33.6~$\pm$~4.1	&	&			\\
	&	Control	&	38.6~$\pm$~6.0	&	&			\\
\sdssr	&	All	&	34.8~$\pm$~4.9	&	&			\\
	&	Control	&	29.2~$\pm$~4.6	&	&			\\
---	&	All Ratio	&	0.97~$\pm$~0.18	&	44.2~$\pm$~14.4	& 21.6\\
	&	Control Ratio	&	1.32~$\pm$~0.29	&	26.3~$\pm$~10.0	& \\
\hline
\sdssg	& Southern	&	46.5~$\pm$~5.3	&	&			\\
	&	Control	    &	43.3$\pm$~5.1	&	&			\\
\sdssr	& Southern	&	47.9~$\pm$~6.6	&	&			\\
	&	Control	    &	37.4~$\pm$~5.6	&	&			\\
---	&	Southern Ratio	&	0.97~$\pm$~0.17	&	36.4~$\pm$~13.0	& 24.9\\
	&	Control Ratio	&	1.16~$\pm$~0.22	&	9.8~$\pm$~9.6	& \\
\enddata
\vspace{-0.3cm}
\end{deluxetable}

\section{Discussion}
\label{sec:dis}

Although the primary objective of this work has been to examine the hot dust phenomenon, we have also obtained new information on the local ISM. In order to put the hot dust measurements in their proper context, we first discuss the implications of our observations for the local ISM.

\subsection{Local ISM}

The $p_{\rm green}$:$p_{\rm red}$ ratio obtained here for the controls represents the first information on the polarimetric colour of the local ISM. Until very recently the level of polarization within the Local Hot Bubble (the cavity in the ISM within which the solar system resides, $\sim$100~pc in size) was below the threshold of accurate measurement. However the development of parts-per-million polarimeters has now resulted in two polarimetric surveys of bright stars within 100 pc. A northern hemisphere survey with the PlanetPol instrument \citep{Bailey2010} is presented along with a southern hemisphere survey by HIPPI \citep{Cotton2016}. The PlanetPol instrument operated in a range from 590 to 1000~nm, having an effective wavelength for a F0 star of 753.8~nm \citep{Hough2006}. The HIPPI survey used the \sdssg~filter -- an effective wavelength of 466.2~nm for a F0 star. The HIPPI survey produced systematically higher polarizations than the one using PlanetPol, but because there was no overlap between the surveys, it was not possible to determine what portion could be ascribed to polarimetric colour and what to differing levels of interstellar polarization in different regions of the sky.

The empirical wavelength dependence of interstellar polarization is given by the Serkowski Law (Equation \ref{eq_serkowski}) \citep{Serkowski1975,Wilking1982}. A typical value for $\lambda_{max}$ is 550~nm \citep{Serkowski1975}, but a wide range of extremes have been reported, e.g. 360~nm to 890~nm \citep{Wilking1982}. The reddest values of $\lambda_{max}$ are associated with dusty nebulae and larger grain sizes \citep[][and references therein]{Clarke2010}. However, all previous work corresponds to regions beyond the Local Hot Bubble. The Local Hot Bubble is a region largely devoid of dust and gas. The results presented here, albeit with large uncertainty, suggest a particularly blue $\lambda_{max}$ for this region; this implies small grain sizes. A dedicated multi-band polarimetric study of stars within the Local Hot Bubble is needed to confirm this result, but it seems clear that polarization in the \sdssg~filter is greater than that in the PlanetPol waveband. However, even allowing for the correction, as can be seen from Fig. \ref{fig:sky_plot} polarization with distance is greater in the south than the north. In fact, had our determination of $\lambda_{max}$ been more ordinary (redder $\lambda_{max}$), the conversion factor applied to the PlanetPol data would have been less, and the difference between the hemispheres starker. A finding of greater interstellar polarization in the south is supported by the survey of \citet{Tinbergen1982} whose was the most sensitive (60 ppm precision) prior to that of \citet{Bailey2010} and \citet{Cotton2016}. He tentatively (using 2-sigma results) identified a ``patch'' of interstellar polarization roughly 30 degrees in angular extent at southern galactic latitudes between 0 and 20 pc, which on an equatorial co-ordinate system corresponds predominantly to southern latitudes.

The results of the HIPPI survey suggested that within the Local Hot Bubble $\sim$2~ppm~pc$^{-1}$ was close to the maximum polarization with distance for any region on the sky in \sdssg. That extreme corresponded predominantly to regions of the southern sky within 30~pc. For the most part the ISM displayed the level of polarization expected. The controls, which were mostly nearby in the south, averaged to 1.81~ppm~pc$^{-1}$ in \sdssg. However, the very high polarization we measure here for nearby star HD~7693 at 7.42~ppm~pc$^{-1}$, challenges this conclusion. HD~7693 is particularly polarimetrically red. A second control, HD~4308, and the hot dust star these two controls are paired with, HD~7788 (which also has a high polarization), also display a similar $p_{\rm green}$:$p_{\rm red}$ ratio. This gives us confidence the ISM is responsible for the higher polarization observed in this region of the sky. Additionally the polarization angles for all three are very similar -- in both green and red -- indicating that a single interstellar dust cloud is primarily responsible for the elevated polarization recorded in this instance. Recent measurements of the hot Jupiter host star HD~189733 \citep{Bott2016} also return a background polarization magnitude in excess of that expected from the trends identified in \citet{Cotton2016}. These results underline how patchy the local ISM can be.

In the previous section mention was made of the dispersion ($\phi$), and how it was consistent with zero for the F/G control group, but not the A control group (see Table \ref{tab:sample_results}). \citet{Treanor1963} was the first to point out that if a light ray passes through two misaligned dust clouds with different chromatic polarimetric properties then this will rotate the position angle of one wavelength with respect to another. \cite{GehSil1965} have observed this dispersion phenomena in more distant clouds. From the A control group, the two stars with the greatest deviation are both northern stars. A reasonable hypothesis to explain this is therefore that the dust in the south is better aligned, predominantly forming a single cloud, whereas the north may contain a variety of unaligned clouds. We can rule out the chance possibility of oppositely aligned strongly polarizing clouds in the north, since the PlanetPol survey results show a smooth increase in polarization with distance there \citep{Bailey2010}. Thus, given that in the north especially, interstellar polarization within the Local Hot Bubble is very low, misaligned but similarly diffuse clouds are likely to be producing the observed effect. 

\subsection{Hot dust} 

In Section \ref{sssec:no_pol}, we determined that the measured polarization of the hot dust stars is consistent with the expected ISM contribution, given the stellar distances. From this we placed limits on the total brightness of any exo-Asteroid belt around these stars assuming the fractional excesses were produced by scattered light. The predicted values for a scattered light origin of the excesses were much greater than the observed polarization magnitudes.

As an aternative to scattered light, we might assume the excess is produced by thermal emission, or some combination of scattered light and thermal emission. In this case, the relationship between polarization and dust luminosity is given by \begin{equation}\label{eq_therm_est} p = (L_{\rm IR}/L_{\star})f_{\rm pol}A \end{equation} Where the polarization $p$, dust fractional luminosity $L_{\rm IR}/L_{\star}$, polarization fraction $f_{\rm pol}$, and albedo $A$. In this instance we need to know $L_{\rm IR}/L_{\star}$, which we calculate to be 3 per cent, under the assumption of blackbody emission with $T_{\rm dust}~\sim~1500$~K from temperature constraints given by the slope of $H$ and $K$ measurements in \cite{Absil2013} and \cite{Ertel2014}. For micron-sized dust grains the albedoes typically span a range 0.1 to 0.3 for silicacious and icy materials, respectively. Assuming a 1 per cent fractional excess in the $H$ band caused by thermal emission from 1500~K dust at 0.1~au around a sun-like star, the optical scattered fractional brightness ($F_{\rm scat}/F_{\star}$) would be in the range $\sim$~20 to 60~ppm \citep{KenWya2011}, consistent with both the non-detection of significant polarization by HIPPI and the derived expectation value of hot dust polarization (for polarization fractions of 5 to 50 per cent). The observed properties of hot dust are therefore consistent with a thermal emission origin for the excess.

However, most of the stars in our sample are more luminous than the sun as adopted in the above example. This requires the adoption of lower albedoes and/or lower polarization fractions to match the measured polarization magnitudes. Low albedo ($\leq$~0.1), low polarization fraction ($\leq$~5 per cent) dust grains are consistent with the measurements, but these properties (in combination) are atypical for micron-sized debris dust grains. Sub-micron (0.1~$\mu$m) dust grains are predicted to have such qualities, and have been attributed as the cause of the hot dust phenomenon through thermal, rather than scattered light, emission \citep{Rieke2015}. Identifying a combination of dust optical properties and material consistent with these observations is beyond the scope of this work, but the parameter space defined here is consistent, in varying parts, with commonly adopted materials. We can rule out scattered light from typical (micron-sized) dust grains as the origin of the near infrared excess, being too bright and strongly polarizing, with the caveat that the geometry of the dust might conspire to mask the total polarization signal.

The F/G hot dust stars have a dispersion inconsistent with zero at a significant level, whilst the controls are consistent with zero. If the southern A star control-target pair is added to the three F/G stars this result is little changed. The $p_{\rm green}$:$p_{\rm red}$ ratio is greater in the hot dust groups than their controls; though not statistically significant, this is likely due to fairly large proportional errors. In combination with the dispersion the difference in ratios hints at an intrinsic polarization by hot dust with a different spectral slope to that of the local ISM. This must be a small effect, much smaller in fact than for more typical debris disk host stars (Marshall in prep.), and suggests that the contribution of the hot dust to the total polarization is smaller than that of the ISM, even for stars as close as these. This is consistent with the hot dust phenomenon being attributed to the presence of small nano-scale grains, which would by nature be weakly polarizing.

As alluded to in the introduction, there are potential mechanisms for inducing low level polarization in stars that do not require circumstellar dust. Be star mechanisms can be ruled out as a cause for stellar polarization in this sample due to the range of spectral types. Similarly, none of the stars are noted for strong stellar activity or photometric hotspots, two further causes polarization. Oblateness has been proposed as a cause of stellar polarization \citep{Ohman1946} that would also have an increasing effect toward longer wavelengths, although this has not been detected. Three of the stars in this sample have been measured, or are postulated, to be oblate to varying degrees -- HD~2262 (0.15), HD~7788 (0.20), and HD~187642 (0.09) \citep{vanBelle2012}. The best candidate for detecting oblateness-induced polarization is Regulus, with a magnitude of $\sim$~37~ppm \citep{Bailey2010}, which rotates at 86 per cent of its critical velocity and has a B8~IV spectral type (Cotton, priv. comm.). The hot dust stars in this sample are all slower rotators and cooler than Regulus, such that the magntitude of the induced polarization should be much smaller \citep{Sonne1982} and it cannot therefore be called upon as a mechanism to explain the observations. A final possibility for interference with the polarization signal of the hot dust phenomenon is the presence of a close sub-stellar companion to the target star, i.e. a hot Jupiter. The predicted amplitude of hot Jupiter-induced polarization is at the 10s ppm level \citep{Seager2000}. The most recent measurements of HD~189733 support a signal of this amplitude \citep{Wiktorowicz2015,Bott2016}.

However, there are scenarios that would result in a non-detection of polarization. Concerning geometry, if the debris was distributed smoothly in a disk oriented face-on, the polarization from the dust would cancel evenly leaving little to no detectable signal; we discount this possibility as being highly unlikely for all six targets in the sample. Likewise a spherical shell of dust grains around the star, perhaps delivered by exo-Oort cloud comets, would also produce little measurable polarization; this is a scenario we cannot test with these observations. Alternatively, time variability of the hot dust may also result in a non-detection if the level of dust was low at the epoch of observation. \cite{Ertel2016} validated previous detection of hot dust stars with VLTI/PIONIER and identified only one case, that of HD~7788 (one of the targets examined here), where the hot excess was detected and found to be variable. The persistence of the hot dust phenomenon over multi-year timescales makes it unlikely that the lack of detections here can be ascribed to variability.

\section{Conclusions}
\label{sec:con}

We have measured the optical polarization of starlight at the parts-per-million level for six hot dust excess stars in two wavebands. These stars do not exhibit excess emission at other, longer wavelengths allowing us to rule out contributions from cooler circumstellar matter to the total polarization. We also observed a number of non-excess control stars closely situated to the target stars in order to characterise the ISM contribution to the total polarization. 

The magnitudes of polarization of the hot dust stars are consistent with those of the ISM controls. Using simple arguments, our observations suggest dust grains with low albedos and low polarization as the origin for the observed excesses, incompatible with the scattered light properties of known circumstellar debris disks. From this we can rule out scattered light from dust in exo-Asteroid belts being responsible for the hot dust excesses. Our results favour the interpretation of hot dust as being due to the thermal emission of nano-scale dust grains trapped in the vicinity of the host star. Whilst a face-on geometry for the discs would produce little measurable polarization, such a geometry would be unusual for all six hot dust stars examined here. We cannot rule out a spherical distribution of dust either, but that is not a scenario we wish to examine here. 

Significant dispersion in the hot dust stars in contrast to southern control stars indicates an intrinsic polarization by hot dust with a different spectral slope to the local ISM. The data implies a greater contribution at redder wavelengths. The contribution of the hot dust to the measured polarization can only be around 30~ppm or less, consistent with small, sub-micron dust grains. 

These observations constitute the first multi-wavelength measurements of the ISM within the Local Hot Bubble, and suggest a particularly blue polarimetric colour for the ISM in this region. In the south, the absence of significant dispersion is consistent with a homogeneous single cloud for the ISM. In general we find levels of interstellar polarization consistent with that found by the HIPPI bright star survey ($\leq~$2~ppm~pc$^{-1}$). A region of the ISM in the direction of HD~7693 is identified where the polarization reaches $>$7~ppm~pc$^{-1}$ illustrating the clumpy nature of the ISM, even within the Local Hot Bubble. 

We have demonstrated the potential of modern aperture polarimeters with their greatly improved precision to probe the inner regions of nearby star systems. No other technique is able to provide the insights into hot dust systems developed here. To make best use of this technique though requires an improved knowledge of the properties of the local ISM, in particular the distribution of dust and its polarimetric colour. We urge multi-band polarimetric mapping of the local ISM as a priority to facilitate further insights.

\section{Acknowledgements}

The authors thank the anonymous referee for their comments which helped improve the manuscript. JPM is supported by a UNSW Vice-Chancellor's postdoctoral fellowship. GMK is supported by the Royal Society as a Royal Society University Research Fellow. This work has been supported by Mexican CONACyT research grant CB-2012-183007 (CdB). This work was supported by the European Union through ERC grant number 279973 (GMK and MCW). This work has made use of observations taken on the Anglo-Australian Telescope. This research made use of the the SIMBAD database and VizieR catalogue access tools, operated at CDS, Strasbourg, France. This research has made use of NASA’s Astrophysics Data System Bibliographic Services. We acknowledge the assistance of Behrooz Karamiqucham in helping take the observations during 29/10-3/11/2015.

\end{document}